\documentclass[twoside]{article}
\usepackage{fleqn,espcrc2}
\usepackage{here}%pour inserer figures et tables a emplacement
\newcommand{\be}{\begin{equation}}
\newcommand{\ee}{\end{equation}}
\newcommand{\bea}{\begin{eqnarray}}
\newcommand{\eea}{\end{eqnarray}}

\def\beq{\begin{equation}}
\def\eeq{\end{equation}}
\def\bea{\begin{eqnarray}}
\def\eea{\end{eqnarray}}
\def\bq{\begin{quote}}
\def\eq{\end{quote}}

\def\nnb{\nonumber}
\def\ga{\left(}
\def\dr{\right)}
\def\aga{\left\{}
\def\adr{\right\}}

\def\rar{\rightarrow}
\def\lrar{\Longrightarrow}

\def\nnb{\nonumber}
\def\la{\langle}
\def\ra{\rangle}
\def\nin{\noindent}
\def\ba{\begin{array}}
\def\ea{\end{array}}

\def\als{\alpha_s}

\def\g2{ \la\alpha_s G^2 \ra}
\def\g3{g^3f_{abc}\la G^aG^bG^c \ra}
\def\g4{\la\als^2G^4\ra}

\newcommand{\AmS}{{\protect\the\textfont2
  A\kern-.1667em\lower.5ex\hbox{M}\kern-.125emS}}

\title{\bf{
$V$-$A$ hadronic
tau decays : a QCD laboratory} \thanks{Talk given at the QCD 00
International Euroconference (Montpellier 6-13th July 2000). This is a summary of the paper
``New QCD Estimate of the Kaon Penguin Matrix Elements" hep-ph/0004247 (Nucl. Phys. B in
press).}} 
\author{
Stephan Narison\address{Laboratoire de Physique Math\'ematique et Th\'eorique,
Universit\'e de Montpellier II
Place Eug\`ene Bataillon,
34095 - Montpellier Cedex 05, France \\ 
E-mail:
narison@lpm.univ-montp2.fr}
}
\begin{document}
\pagestyle{empty}
\begin{abstract}
\nin
Recent ALEPH/OPAL data on the $V$-$A$ spectral functions from hadronic $\tau$ decays are
used for fixing the QCD continuum threshold
at which the first and second Weinberg sum rules should be satisfied in the
chiral limit, and for predicting the values
of the low-energy constants $f_\pi,~m_{\pi^+}-m_{\pi^0}$ and ${ L}_{10}$.
Some DMO-like sum rules and the $\tau$-total hadronic widths
$R_{\tau,V-A}$ are also used for extracting
the values of the $D=6,~8$ QCD vacuum condensates and the corresponding (in the
chiral limit)  electroweak kaon penguin matrix elements
$\la {\cal Q}^{3/2}_{8,7}\ra_{2\pi}$, , where a deviation from the vacuum saturation estimate has been
obtained. Combining these results with the one of the QCD penguin matrix element
$\la {\cal Q}^{1/2}_{6}\ra_{2\pi}$ obtained from a (maximal) $\bar
qq$-gluonium mixing scheme from the scalar meson sum rules, we deduce, in the
Electroweak Standard Model (ESM), the conservative upper bound  for the CP-violating ratio:
$\epsilon'/\epsilon \leq (22\pm 9) 10^{-4}$, in agreement with the present measurements.
\end{abstract}
\maketitle
\textheight 24.5cm
\topmargin -2.cm
%\oddsidemargin +0.2cm
%\evensidemargin -1.0cm
%\maketitle
\section{Introduction}
\nin
Hadronic tau decays have been demonstrated \cite{BNP}  (hereafter referred as BNP) to be an
efficient laboratory for testing perturbative and non-perturbative QCD. That is due both to the
exceptional value of the tau mass situated at a frontier regime between
perturbative and non-perturbative QCD and to the excellent quality of the
ALEPH/OPAL \cite{ALEPH,OPAL} data. On the other, it is also known before the
advent of QCD, that the Weinberg \cite{WEINBERG} and DMO \cite{DMO} sum rules
are important tools for controlling the chiral and flavour symmetry realizations
of QCD, which are broken by light quark mass terms to higher order \cite{FNR}
and by higher dimensions QCD condensates \cite{SNWEIN} within the SVZ expansion
\cite{SVZ}. \\ In this talk, we
shall discuss the impact of the new ALEPH/OPAL data on the $V$-$A$ spectral functions
in the analysis of the previous and some other related sum rules, which will be used
for determining the low-energy constants of the effective chiral lagrangian \cite{CHPT,RAFAEL,CHPT2},
the SVZ QCD vacuum condensates \cite{SVZ}. In
particular, we shall discuss the consequences of these results on the estimate of the kaon 
$CP$--violation parameter $\epsilon'/\epsilon$ in the Electroweak Standard Model
(ESM). These results have been originally obtained in \cite{CP} and will be reviewed here.
\section{Tests of the ``sacrosante"  Weinberg and DMO sum rules in the chiral limit}
\subsection{Notations}
\nin
We shall be concerned here with the two-point correlator:
\bea
\Pi^{\mu\nu}_{LR}(q)&\equiv& i\int d^4 x~ e^{iqx}\la 0|{\cal T} J^\mu_L(x)\ga
J^\nu_R(0)\dr^{\dagger}|0\ra\nnb\\
&=&-(g^{\mu\nu}q^2-q^\mu q^\nu)\Pi_{LR}(q^2)~,
\eea
built from the left-- and right--handed components of the local weak current:
\beq
J^\mu_{L}=\bar u\gamma^\mu(1-\gamma_5)d,~~~~~~~J^\mu_{R}=\bar
u\gamma^\mu(1+\gamma_5)d~,
\eeq
and/or using isospin rotation relating the neutral and charged weak currents:
\beq
\rho_V-\rho_A\equiv \frac{1}{2\pi}{\rm Im}\Pi_{LR}\equiv\frac{1}{4\pi^2}\ga
v-a\dr~.
\eeq
The first term is the notation in \cite{DONO}, while the last one is the
notation in \cite{ALEPH,OPAL}.
\subsection{The sum rules}
\nin
The ``sacrosante" DMO and Weinberg sum rules read in the chiral limit
\footnote{Systematic analysis of the breaking of these sum rules by light
quark masses \cite{FNR} and
condensates \cite{SNWEIN,SVZ} within the context of QCD have been done
earlier.}:
\bea
{\cal S}_{0}&\equiv&\int_0^{\infty} ds ~\frac{1}{2\pi}{\rm Im}\Pi_{LR} =
f^2_\pi~,\nnb\\
{\cal S}_{1}&\equiv&\int_0^{\infty} ds ~s~\frac{1}{2\pi}{\rm Im}\Pi_{LR} =
0~,\nnb\\
{\cal S}_{-1}&\equiv&\int_0^{\infty} \frac{ds}{s}~ \frac{1}{2\pi}{\rm
Im}\Pi_{LR} = -4L_{10}~,\nnb\\
{\cal S}_{em}&\equiv&\int_0^{\infty} ds~\ga s~\log\frac{s}{\mu^2}\dr
\frac{1}{2\pi}{\rm Im}\Pi_{LR} \nnb\\&=&
-\frac{4\pi}{3\alpha}f^2_\pi \ga m^2_{\pi^\pm}-m^2_{\pi^0}\dr~,
\eea
where $f_\pi\vert_{exp}=(92.4\pm 0.26)$ MeV is the experimental pion decay
constant which should be used
here as we shall use data from $\tau$-decays involving physical pions;
$m_{\pi^\pm}-m_{\pi^0}\vert_{exp}\simeq 4.5936(5)$ MeV; $L_{10}\equiv
f_\pi^2{\la r_\pi^2\ra}/{3}-F_A$
{[}$\la r_\pi^2\ra=(0.439\pm 0.008)fm^2$ is the mean pion radius and
$F_A=0.0058\pm 0.0008$ is the
axial-vector pion form factor for $\pi\rar e\nu\gamma${]} is one the
low-energy constants of the effective
chiral lagrangian
\cite{CHPT,RAFAEL,CHPT2}. In order to exploit these sum rules using the ALEPH/OPAL
\cite{ALEPH,OPAL} data from the hadronic tau--decays, we shall work with
their Finite
Energy Sum Rule (FESR) versions (see e.g. \cite{FNR,BNP} for such a
derivation). In the chiral
limit  ($m_q=0$ and
$\la
\bar uu\ra=\la
\bar dd\ra =\la \bar ss\ra$), this is equivalent to truncate the LHS at
$t_c$ until which the data are available, while the RHS of the integral
remains valid to leading order
in the 1/$t_c$ expansion in
the chiral limit, as the breaking of
these sum rules by higher dimension $D=6$ condensates in the chiral limit
which is of the order of
$1/t_c^3$ is numerically negligible \cite{SNWEIN}.
\subsection{Matching between the low and high-energy regions}
\nin
In order to fix the $t_c$ values which separate the low and high energy
parts of the spectral functions, we
require that the 2nd Weinberg sum rule (WSR) ${\cal S}_1$ should be
satisfied by the present data. 
As shown
in Fig. 1 (see \cite{CP}), this is obtained for two
values of $t_c$ \footnote{One can compare the two solutions with the
$t_c$--stability region around 2 GeV$^2$ in the
QCD spectral sum rules analysis (see e.g. Chapter 6 of \cite{SNB}).}:
\beq\label{weinberg}
t_c\simeq (1.4\sim 1.5)~{\rm GeV}^2~~~{\rm and}~~~
(2.4\sim 2.6)~{\rm GeV}^2.
\eeq
\begin{figure}[H]
\begin{center}
\caption{ FESR version of the 2nd Weinberg sum rule versus $t_c$ in GeV$^2$ using the
ALEPH/OPAL data of the spectral functions. Only the central values are shown.}
\end{center}
\end{figure}
\nin
Though the 2nd value is interesting from the point of view of the QCD
perturbative calculations (better convergence of the QCD series), its exact
value is strongly affected by the
inaccuracy of the data near the
$\tau$--mass (with the low values of the ALEPH/OPAL data points, the 2nd
Weinberg sum rule is only
satisfied at the former value of $t_c$).\\
After having these $t_c$ solutions, we can improve the constraints by
requiring that the 1st Weinberg sum rule ${\cal
S}_0$ reproduces the experimental value of
$f_\pi$ \footnote{Though we are working here in the chiral limit, the data
are obtained for physical pions,
such that the corresponding value of $f_\pi$ should also correspond to the
experimental one.}
within an accuracy 2-times the experimental error. 
This condition allows to fix
$t_c$ in a very narrow margin due to the sensitivity of the result on the
changes of $t_c$ values
\footnote{For the
second set of
$t_c$-values in Eq. \ref{weinberg}, one obtains a slightly lower value:
$f_\pi=(84.1\pm 4.4)$ MeV.}
\beq\label{tc}
t_c=(1.475\pm 0.015)~{\rm GeV}^2~,
\eeq
\section{
Low-energy constants $L_{10}$, $m_{\pi^\pm}-m_{\pi^0}$ and $f_\pi$ in the chiral limit
}
\nin
Using the previous value of $t_c$ into the ${\cal S}_{-1}$ sum rule, we deduce:
\beq\label{l10}
L_{10}\simeq -(6.26\pm 0.04)\times 10^{-3}~,
\eeq
which agrees quite well with more involved analysis including chiral
symmetry breakings \cite{STERN,OPAL},
and with the one using a lowest meson dominance (LMD) of the spectral
integral \cite{ENJL}. \\
Analogously, one obtains from the ${\cal S}_{em}$ sum rule:
\beq\label{deltam}
\Delta m_\pi\equiv m_{\pi^\pm}-m_{\pi^0}\simeq (4.84\pm 0.21)~{\rm MeV}~.
\eeq
This result is 1$\sigma$ higher than the data $4.5936(5)$ MeV, but agrees
within the errors with the more detailed
analysis from $\tau$--decays \cite{PECCEI,OPAL} and with the LMD result of
about 5 MeV \cite{ENJL}. We have checked
that moving the subtraction point $\mu$ from 2 to 4 GeV slightly decreases
the value of
$\Delta m_\pi$ by $3.7\%$ which is relatively weak, as expected. Indeed, in
the chiral limit, the $\mu$
dependence does not appear in the RHS of the ${\cal S}_{em}$ sum rule, and
then, it looks
natural to choose:
\beq
\mu^2=t_c~,
\eeq
because $t_c$ is the only external scale in the analysis. At this scale the
result increases
slightly by 2.5\%. One can also notice that the prediction for $\Delta m$
is more stable when one changes the
value of $t_c=\mu^2$. Therefore, the final predictions from the value of
$t_c$ in Eq. (\ref{tc}) fixed from the 1st
and 2nd Weinberg sum rules are:
\bea\label{pred}
\Delta m &\simeq& (4.96\pm 0.22)~{\rm MeV}~,\nnb\\ L_{10}&\simeq&-(6.42\pm
0.04)\times 10^{-3}~,
\eea
which we consider as our "best" predictions. \\
For some more conservative results, we also give the predictions obtained from
the second $t_c$--value given in Eq. (\ref{weinberg}). In this way, one
obtains:
\bea
f_\pi&=&(87\pm 4)~{\rm MeV}~,\nnb\\ \Delta m &\simeq& (3.4\pm 0.3)~{\rm
MeV}~,\nnb\\ L_{10}&\simeq& -(5.91\pm
0.08)\times 10^{-3}~,
\eea
where one can notice that the results are systematically lower
than the ones obtained in Eq. (\ref{pred}) from the first $t_c$--value
given previously, which may disfavour
a posteriori the second choice of $t_c$-values, though we do not have a
strong argument favouring one with respect
to the other \footnote{Approach based on
$1/N_c$ expansion and a saturation of the spectral function by the lowest
state within a narrow width
approximation (NWA) favours the former value of $t_c$ given in Eq.
(\ref{tc}) \cite{ENJL}.}. 
Therefore, we take as a conservative value the largest range
spanned by the two sets of results,
namely:
\bea\label{conserve}
f_\pi&=&(86.8\pm 7.1)~{\rm MeV}~,\nnb\\ \Delta m &\simeq& (4.1\pm 0.9)~{\rm
MeV}~,\nnb\\ L_{10}&\simeq& -(5.8\pm
0.2)\times 10^{-3}~,
\eea
which we found to be quite satisfactory in the chiral limit.
The previous tests are very useful, as they will allow us to jauge the
confidence level of the next predictions.
\section{Soft pion and kaon reductions of $\la (\pi\pi)_{I=2}|{\cal
Q}^{3/2}_{7,8}|K^0\ra$ to vacuum condensates}
\nin
We shall consider here the kaon electroweak penguin matrix elements:
\beq
\la {\cal Q}^{3/2}_{7,8}\ra_{2\pi}\equiv \la (\pi\pi)_{I=2}|{\cal
Q}^{3/2}_{7,8}|K^0\ra~,
\eeq
defined as:
\bea
{\cal Q}_7&\equiv& \frac{3}{2}\ga \bar s d\dr_{V-A}\sum_{u,d,s}e_\psi\ga \bar
\psi\psi\dr_{V+A}~,\nnb\\ {\cal Q}_8&\equiv&
\frac{3}{2}\ga \bar s_\alpha
d_\beta\dr_{V-A}\sum_{u,d,s}e_\psi\ga \bar
\psi_\beta\psi_\alpha\dr_{V+A}~,
\eea
where $\alpha,\beta$ are colour indices; $e_\psi$ denotes the electric charges.
In the chiral limit $m_{u,d,s}\sim m^2_\pi\simeq m^2_K=0$, one can use soft
pion and kaon techniques in order
to relate the previous amplitude to the four-quark vacuum condensates
\cite{DONO} (see also \cite{ENJL}):
\bea\label{soft}
\la {\cal Q}^{3/2}_{7}\ra_{2\pi}&\simeq &-\frac{4}{f_\pi^3}\la
{\cal O}^{3/2}_7\ra~,\nnb\\
\la {\cal Q}^{3/2}_{8}\ra_{2\pi}&\simeq
&-\frac{4}{f_\pi^3}\aga\frac{1}{3}\la {\cal O}^{3/2}_7\ra+\frac{1}{2}\la
{\cal O}^{3/2}_8\ra\adr~,
\eea
where we use the shorthand notations: $\la 0|{\cal O}^{3/2}_{7,8}|0\ra\equiv
\la {\cal O}^{3/2}_{7,8}\ra $, and $f_\pi=(92.42\pm 0.26)$ MeV \footnote{In
the chiral limit $f_\pi$ would be
about 84 MeV. However, it is not clear to us what value of $f_\pi$ should
be used here, so we shall leave it as
a free parameter which the reader can fix at his convenience.}. Here:
\bea
{\cal O}^{3/2}_7&=&\sum_{u,d,s}
\bar\psi\gamma_\mu\frac{\tau_3}{2}\psi\bar\psi\gamma_\mu\frac{\tau_3}{2}\psi\nnb\\&&-
\bar\psi\gamma_\mu\gamma_5\frac{\tau_3}{2}\psi\bar\psi\gamma_\mu\gamma_5\frac{
\tau_3}{2}\psi~,\nnb\\
{\cal O}^{3/2}_8&=&\sum_{u,d,s}
\bar\psi\gamma_\mu\lambda_a\frac{\tau_3}{2}\psi\bar\psi\gamma_\mu\lambda_a\frac{
\tau_3}{2}\psi\nnb\\&&-
\bar\psi\gamma_\mu\gamma_5\lambda_a\frac{\tau_3}{2}\psi\bar\psi\gamma_\mu\gamma_
5\lambda_a\frac{\tau_3}{2}\psi~,
\eea
where $\tau_3$ and $\lambda_a$ are flavour and colour matrices.
Using further pion and kaon reductions in
the chiral limit, one can relate this matrix element to the
$B$-parameters:
\bea\label{soft2}
B^{3/2}_7&\simeq& \frac{3}{4}\frac{\ga m_u+m_d\dr
}{m^2_\pi}\frac{\ga m_u+m_s\dr
}{m^2_K}\frac{1}{f_\pi}
\la {\cal Q}^{3/2}_{7}\ra_{2\pi}\nnb\\
B^{3/2}_8&\simeq& \frac{1}{4}\frac{\ga m_u+m_d\dr
}{m^2_\pi}\frac{\ga m_u+m_s\dr
}{m^2_K}\frac{1}{f_\pi}
\la {\cal Q}^{3/2}_{8}\ra_{2\pi}\nnb\\
\eea
where all QCD quantities will be evaluated in the $\overline{MS}$-scheme
and at the scale $M_\tau$.
\section{ The $
\la {\cal O}^{3/2}_{7,8}\ra $ vacuum condensates from DMO-like sum rules
in the chiral limit}
\nin
In previous papers \cite{DONO,ENJL}, the vacuum condensates 
$\la {\cal O}^{3/2}_{7,8}\ra $ have been extracted
using Das-Mathur-Okubo(DMO)-- and Weinberg--like sum rules based on the
difference of the vector and axial-vector
spectral functions $\rho_{V,A}$  of the
$I=1$ component of the neutral current:
\bea\label{dispersive}
2\pi\la \alpha_s {\cal O}^{3/2}_8\ra&=&\int_0^\infty ds
~s^2\frac{\mu^2}{s+\mu^2}\ga
\rho_V-\rho_A\dr~,\nnb\\
\frac{16\pi^2}{3}\la {\cal O}^{3/2}_7\ra &=&\int_0^\infty ds~
s^2\times\nnb\\
&&\log\ga\frac{s+\mu^2}{s}\dr\ga
\rho_V-\rho_A\dr~,
\eea
where $\mu$ is the subtraction point.
Due to the quadratic divergence of the integrand, the previous
sum rules are expected to be sensitive to the high energy tails of the
spectral functions where the
present ALEPH/OPAL data from $\tau$-decay \cite{ALEPH,OPAL} are inaccurate.
This inaccuracy can a
priori affect the estimate of the four-quark vacuum condensates. On the
other hand,
the explicit
$\mu$--dependence of the analysis can also induce another uncertainty.
En passant, we check below the effects of these two parameters $t_c$ and $\mu$.
After evaluating the spectral integrals, we obtain at $\mu$= 2 GeV and for
our previous values of $t_c$ in Eq. (\ref{tc}), the values (in units of
$10^{-3}$ GeV$^6$)
using the cut-off momentum scheme (c.o):
\bea\label{mom}
\alpha_s\la{\cal O}^{3/2}_{8}\ra_{c.o}&\simeq& -(0.69\pm 0.06)~,\nnb\\
\la{\cal O}^{3/2}_{7}\ra_{c.o}&\simeq& -(0.11\pm 0.01)~,
\eea
where the errors come mainly from the small changes of $t_c$--values. If
instead, we use the second set of values
of $t_c$ in Eq. (\ref{weinberg}), we obtain by setting $\mu$=2 GeV:
\bea\label{mom2}
\alpha_s\la{\cal O}^{3/2}_{8}\ra_{c.o}&\simeq& -(0.6\pm 0.3)~,\nnb\\
\la{\cal O}^{3/2}_{7}\ra_{c.0}&\simeq& -(0.10\pm 0.03)~,
\eea
which is consistent with the one in Eq. (\ref{mom}), but with larger errors
as expected. We have also checked that
both $\la{\cal O}^{3/2}_{8}\ra$ and $\la{\cal O}^{3/2}_{7}\ra$ increase in
absolute value when $\mu$ increases where
a stronger change is obtained for $\la{\cal O}^{3/2}_{7}\ra$, a feature
which has been already noticed in
\cite{ENJL}. In order to give a more conservative estimate, we consider as
our final value the largest range
spanned by our results from the two different sets of $t_c$--values. This
corresponds to the one in Eq. (\ref{mom2})
which is the less accurate prediction.
 We shall
use the relation between the momentum cut-off (c.o) and
$\overline{MS}$--schemes given in
\cite{DONO}:
\bea\label{rel}
\la{\cal O}^{3/2}_{7}\ra_{\overline{MS}}&\simeq& \la{\cal O}^{3/2}_{7}\ra_{c.o}
\nnb\\ &&+ \frac{3}{8}a_s\ga \frac{3}{2}+2d_s\dr\la{\cal O}^{3/2}_{8}\ra\nnb\\
\la{\cal O}^{3/2}_{8}\ra_{\overline{MS}}&\simeq& \ga 1-\frac{119}{24}a_s
\pm \ga\frac{119}{24}a_s\dr^2\dr
\nnb\\ &&\times \la{\cal O}^{3/2}_{8}\ra_{c.o}
-a_s\la{\cal O}^{3/2}_{7}\ra~,
\eea
where $d_s=-5/6$ (resp 1/6) in the so-called Na\"\i ive Dimensional Regularization NDR (resp.
t'Hooft-Veltmann HV) schemes \footnote{The two schemes differ by the treatment of the $\gamma_5$
matrix.};
$a_s\equiv
\alpha_s/\pi$. One can notice that the $a_s$ coefficient is large in the
2nd relation (50\%
correction), and the situation is worse because of the
relative minus sign
between the two contributions. Therefore, we have added a rough estimate of
the $a_s^2$ corrections based on the na\"\i ve growth of the PT series,
which here gives 50\%
corrections of the sum of the two first terms. For
a consistency of the whole approach, we shall use the value of
$\alpha_s$ obtained from
$\tau$--decay, which is \cite{ALEPH,OPAL}:
\bea
\alpha_s(M_\tau)|_{exp}&=& 0.341\pm 0.05~\lrar\nnb\\
\alpha_s(2~{\rm GeV})&\simeq&
0.321\pm 0.05~.
\eea
Then, we deduce (in units of $10^{-4}$ GeV$^6$) at 2 GeV:
\bea\label{dono78}
\la{\cal O}^{3/2}_{7}\ra_{\overline{MS}}&\simeq& -(0.7\pm 0.2)~,\nnb\\
\la{\cal O}^{3/2}_{8}\ra_{\overline{MS}}&\simeq&-(9.1\pm 6.4)~,
\eea
where the large error in $\la{\cal O}^{3/2}_{8}\ra$ comes from the estimate
of the $a_s^2$ corrections
appearing in Eq. (\ref{rel}). In terms of the $B$ factor and with the
previous value of
the light quark masses in Eq. (\ref{ms}), this result, at $\mu=2$ GeV, can be
translated into:
\bea\label{b78}
B^{3/2}_{7}&\simeq& (0.7\pm 0.2)~\ga\frac{m_s(2)~[{\rm
MeV}]}{119}\dr^2k^4 ~,\nnb\\
B^{3/2}_{8}&\simeq&(2.5\pm 1.3)~\ga\frac{m_s(2)~[{\rm
MeV}]}{119}\dr^2k^4~.
\eea
where: \beq\label{k}
k\equiv \frac {92.4}{f_\pi~[{\rm MeV}]}~.
\eeq
\begin{itemize}
\item
Our results in Eqs. (\ref{dono78}) compare quite well with the ones obtained by
\cite{DONO} in the $\overline{MS}$--scheme (in units of $10^{-4}$ GeV$^6$)
at 2 GeV:
\bea
\la{\cal O}^{3/2}_{8}\ra_{\overline{MS}}&\simeq& -(6.7\pm 0.9)~,\nnb\\
\la{\cal O}^{3/2}_{7}\ra_{\overline{MS}}&\simeq& -(0.70\pm 0.10)~,
\eea
using the same sum rules but presumably a slightly different method for the
uses of the data and for the choice
of the cut-off in the evaluation of the spectral integral.
\item Our errors in the evaluation of the spectral integrals, leading to
the values in Eqs.
(\ref{mom}) and (\ref{mom2}), are mainly due to the slight change of the
cut-off value $t_c$ \footnote{A slight
deviation from such a value affects notably previous predictions as the
$t_c$-stability of the results ($t_c\approx 2$ GeV$^2$) does not coincide
with the one required by the 2nd
Weinberg sum rules. At the stability point the predictions are about a
factor 3 higher than the one obtained
previously.}.
\item The error due to the passage into the ${\overline{MS}}$--scheme is
due mainly
to the truncation of the QCD series, and is important (50\%) for $\la{\cal
O}^{3/2}_{8}\ra$ and $B^{3/2}_8$,
which is the main source of errors in our estimate.
\item As noticed earlier, in the analysis of the pion mass-difference, it
looks more natural to do the
subtraction at $t_c$. We also found that moving the value of
$\mu$ can affects the value of $B^{3/2}_{7,8}$ .
\end{itemize}
For the above reasons, we expect that the results
given in
\cite{DONO} for $\la{\cal O}^{3/2}_{8}\ra$ though interesting are quite
fragile, while the errors quoted there
have been presumably underestimated. Therefore, we think that a
reconsideration of these results using
alternative methods are mandatory.
\section{The $
\la {\cal O}^{3/2}_{7,8}\ra $ vacuum condensates from the hadronic tau total
decay rates}
\nin
In the following, we shall
not introduce any new sum rule, but, instead, we shall exploit known
informations from the
total $\tau$--decay rate and available results from it, which have not the
previous drawbacks. The $V$-$A$ total
$\tau$--decay rate, for the $I=1$ hadronic component, can be deduced from
BNP \cite{BNP},
and reads \footnote{Hereafter we shall work in the $\overline{MS}$--scheme.}:
\beq\label{rate}
{ R}_{\tau,V-A}=\frac{3}{2}\vert V_{ud}\vert
^2S_{EW}\sum_{D=2,4,...}{\delta^{(D)}_{V-A}}~.
\eeq
$\vert V_{ud}\vert =0.9753\pm 0.0006$ is the CKM-mixing angle, while
$S_{EW}=1.0194$ is the electroweak corrections
\cite{MARC}. In the following, we shall use the BNP results for ${\cal
R}_{\tau,V/A}$ in order to deduce ${
R}_{\tau,V-A}$:
\begin{itemize}
\item The chiral invariant
$D=2$ term due to a short distance tachyonic gluon mass \cite{ZAK,CNZ}
cancels in the $V$-$A$ combination.
 Therefore, the $D=2$ contributions come only from the quark mass terms:
\beq
M^2_\tau\delta^{(2)}_{V-A}\simeq 8\Big{[}
1+\frac{25}{3}a_s(M_\tau)\Big{]}m_um_d~,
\eeq
as can be obtained from the first calculation \cite{FNR}, where
$m_u\equiv m_u(M_\tau)\simeq (3.5\pm 0.4)$ MeV,
$m_d\equiv m_d(M_\tau)\simeq (6.3\pm 0.8)$ MeV \cite{SNL}  are respectively
the running coupling and  quark masses evaluated at the scale $M_\tau$.
\item The dimension-four condensate
contribution reads:
\bea
M^4_\tau\delta^{(4)}_{V-A}&\simeq& 32\pi^2\ga 1+\frac{9}{2}a_s^2\dr m^2_\pi
f^2_\pi\nnb\\
&&+{\cal O} \ga m^4_{u,d}\dr~,
\eea
where we have used the $SU(2)$ relation $\la \bar uu\ra=\la \bar dd\ra$ and
the Gell-Mann-Oakes-Renner PCAC
relation:
\beq
(m_u+m_d)\la
\bar uu+\bar dd\ra=-2m^2_\pi f^2_\pi~.
\eeq
\item By inspecting the structure of the combination of dimension-six
condensates entering in
${ R}_{\tau,V/A}$ given by BNP \cite{BNP}, which are renormalizaton group
invariants, and using a $SU(2)$
isospin rotation which relates the charged and neutral (axial)--vector
currents, the $D=6$ contribution reads:
\bea\label{o8}
M^6_\tau\delta^{(6)}_{V-A}&=&-2\times 48\pi^4a_s\Bigg{[}
\Bigg{[} 1+\frac{235}{48}a_s \nnb\\ &&\pm
\ga\frac{235}{48} a_s\dr^2-\frac{\lambda^2}{M^2_\tau}\Bigg{]} \la
{\cal O}^{3/2}_8\ra\nnb\\
&&+a_s\la {\cal O}^{3/2}_7\ra\Bigg{]}~,
\eea
where the overall factor 2 in front expresses the different normalization
between the neutral isovector and
charged currents used respectively in \cite{DONO} and \cite{BNP}, whilst
all quantities are evaluated at the
scale $\mu=M_\tau$. The last two terms in the Wilson coefficients of $\la
{\cal O}^{3/2}_8\ra$ are new: the first term is an estimate of the NNLO term by
assuming a na{\"\i}ve geometric  growth of the $a_s$ series; the second one
is the effect of a tachyonic gluon
mass introduced in
\cite{CNZ}, which takes into account the resummation of the QCD asymptotic
series, with:
$a_s\lambda^2\simeq -0.06$ GeV$^2$
\footnote{This contribution may compete with the dimension-8 operators discussed
in \cite{DONO3}.}. Using
the values of
$\alpha_s(M_\tau)$ given previously, the corresponding QCD series behaves quite well as:
 \bea
{\rm Coef.}~\la{\cal O}^{3/2}_8\ra&\simeq&  1+(0.53\pm 0.08)\nnb\\ &&\pm
0.28+0.18~,
\eea
where the first error comes from the one of $\alpha_s$, while the second
one is due to the unknown
$a_s^2$--term, which introduces an uncertainty of 16\% for the whole
series. The last term is due to the
tachyonic gluon mass.
This leads to the numerical value:
\bea
M^6_\tau\delta^{(6)}_{V-A}&\simeq&-(1.015\pm 0.149)\times 10^3\nnb\\
&&\times\Big{[} (1.71\pm
0.29)\la
{\cal O}^{3/2}_8\ra\nnb\\
&&+a_s\la {\cal O}^{3/2}_7\ra\Big{]}~,
\eea
\item If, one estimates the $D=8$ contribution using a vacuum saturation
assumption, the
relevant $V$-$A$ combination vanishes to leading order of the chiral symmetry
breaking terms. Instead, we
shall use the combined ALEPH/OPAL \cite{ALEPH,OPAL} fit for
$\delta^{(8)}_{V/A}$, and deduce:
\beq
\delta^{(8)}_{V-A}\vert_{exp}=- (1.58\pm 0.12)\times 10^{-2}~.
\eeq
\end{itemize} We shall also use the  combined ALEPH/OPAL data for ${
R}_{\tau,V/A}$, in order to obtain:
\beq
{ R}_{\tau,V-A}\vert_{exp}= (5.0\pm 1.7)\times 10^{-2},~~~~~~~~~~~
\eeq
Using the previous informations into the expression of the rate given in
Eq. (\ref{rate}), one can
deduce:
\beq
\delta^{(6)}_{V-A}\simeq (4.49\pm 1.18)\times 10^{-2}~.
\eeq
This result is in good agreement with the result obtained by using the
ALEPH/OPAL fitted mean value
for $\delta^{(6)}_{V/A}$:
\beq\label{aleph}
\delta^{(6)}_{V-A}\vert_{fit}\simeq (4.80\pm 0.29)\times 10^{-2}~.
\eeq
We shall use as a final result the average of these two determinations,
which coincides with
the most precise one in Eq. (\ref{aleph}). We shall also use the
result:
\beq\label{ratio78}
  \frac{\la{\cal
O}^{3/2}_7\ra}{\la {\cal O}^{3/2}_8\ra}\simeq \frac{1}{8.3}~\ga {\rm
resp.}~ \frac{3}{16}\dr~,
\eeq
where, for the first number we use the value of the ratio of $B^{3/2}_7/
B^{3/2}_8$ which is about $0.7\sim 0.8$ from e.g. lattice calculations
quoted in Table 1, and
the formulae in Eqs. (\ref{soft}) to (\ref{soft2}); for the second number
we use the vacuum saturation
for the four-quark vacuum condensates \cite{SVZ}. The result in Eq.
(\ref{ratio78}) is also comparable with the
estimate of \cite{DONO} from the sum rules given in Eq.(\ref{dispersive}).
Therefore, at the scale
$\mu=M_\tau$, Eqs. (\ref{o8}), (\ref{aleph}) and (\ref{ratio78}) lead, in
the $\overline{MS}$--scheme, to:
\beq\label{resO8}
\la{\cal O}^{3/2}_8\ra\ga M_\tau\dr\simeq -(0.94\pm 0.21)\times
10^{-3}~{\rm GeV}^6~,
\eeq
where the main errors come from the estimate of the unknown higher order
radiative corrections.
It is instructive to compare this result with the one using the vacuum
saturation assumption for the
four-quark condensate (see e.g. BNP):
\bea\label{O8}
\la{\cal O}^{3/2}_8\ra|_{v.s}&\simeq&-\frac{32}{18}\la \bar uu\ra^2\ga
M_\tau\dr\nnb\\ &\simeq& -0.65\times
10^{-3}~\rm{GeV}^6~,
\eea
which shows a $1\sigma$ violation of this assumption. This result is not quite surprising,
as analogous deviations from the vacuum saturation have been already observed in other
channels \cite{SNB}. We have
used for the estimate of
$\la
\bar\psi\psi\ra$the value of $(m_u+m_d)(M_\tau)\simeq 10$ MeV \cite{SNL}
and the GMOR pion PCAC relation. However,
this violation  of the vacuum saturation is
not quite surprising, as a similar fact has also been observed in other
channels
\cite{SNB,ALEPH,OPAL}, though it also appears that the vacuum saturation
gives a quite
good approximate value of the ratio of the condensates
\cite{SNB,ALEPH,OPAL}. The result in Eq. (\ref{resO8})
is comparable with
the value $-(.98\pm 0.26)\times 10^{-3}~{\rm GeV}^6$ at $\mu$=2 GeV
$\approx M_\tau$ obtained by \cite{DONO}
using a DMO--like sum rule, but, as discussed previously, the DMO--like sum
rule result is very sensitive to
the value of $\mu$ if one fixes $t_c$ as in Eq. (\ref{tc}) according to the
criterion discussed above. Here,
the choice $\mu=M_\tau$ is well-defined, and then the result becomes more
accurate (as mentioned
previously our errors come mainly from the estimated unknown $\alpha_s^3$
term of the QCD series).
Using Eqs. (\ref{soft}) and (\ref{ratio78}), our previous result in Eq.
(\ref{resO8}) can be translated into the
prediction on the weak matrix elements in the chiral limit and at the scale $M_\tau$
($k$ is defined in Eq. (\ref{k})):
\bea\label{resq8}
 \la (\pi\pi)_{I=2}|{\cal Q}_{8}^{3/2}|K^0\ra\simeq(2.58\pm
0.58)~{\rm
GeV}^3~k^3\eea
normalized to $f_\pi$, which avoids the ambiguity on the real value of
$f_\pi$ to be used in a such expression.
Our result is higher by about a factor 2 than the quenched
lattice result \cite{MARTI}. A
resolution of this discrepancy can only be done after the inclusion of
chiral corrections in Eqs.
(\ref{soft}) to (\ref{soft2}), and after the uses of dynamical fermions on
the lattice. However, some
parts of the chiral corrections in the estimate of the vacuum condensates
are already included into the QCD expression of the
$\tau$-decay rate and these corrections are negligibly small. We might
expect that chiral corrections,
 which are smooth functions of $m^2_\pi$ will not affect
strongly the relation in Eqs. (\ref{soft}) to (\ref{soft2}), though an
evaluation of their exact size is
mandatory.
Using the previous mean values of the light quark running masses
\cite{SNL}, we deduce in the
chiral limit and at the scale $M_\tau$:
\beq\label{resb8}
B^{3/2}_8\simeq (1.70\pm 0.39)\ga\frac{m_s(M_\tau)~[{\rm
MeV}]}{119}\dr^2k^4,
\eeq
where $k$ is defined in Eq. (\ref{k}). One should notice
that,  contrary to the $B$-factor, the result in Eq. (\ref{resq8}) is
independent to leading order on value of the light quark masses.
\section{Impact of our results on the $CP$-violation parameter
$\epsilon'/\epsilon$}
\nin
One can combine the previous result of $B_8$ with the value of the $B_6$ parameter 
 of
the QCD penguin diagram \cite{BURAS}:
\bea\label{matrix}
\la {\cal Q}^{1/2}_{6}\ra_{2\pi}&\equiv&\la (\pi^+\pi^-)_{I=0}|{\cal
Q}_6^{1/2}|K^0\ra\nnb\\&\simeq&-\Big{[}2\la\pi^+|\bar
u\gamma_5 d|0\ra\la\pi^-|\bar su|K^0\ra+\nnb\\
&&\la \pi^+\pi^-|\bar dd+\bar uu |0\ra\la 0|\bar s\gamma_5
d|K^0\ra~\Big{]}\nnb\\
&\simeq& -4\sqrt{\frac{3}{2}}\ga
\frac{m^2_K}{m_s+m_d}\dr^2\times\nnb\\
&&\sqrt{2}\ga
f_K-f_\pi\dr B^{1/2}_6(m_c)~.
\eea
We have estimated the $\la {\cal
Q}^{1/2}_{6}\ra_{2\pi}$ matrix element by relating its 1st term to the
$K\rar
\pi~ l\nu_l$ semi-leptonic form factors  as usually done (see e.g. \cite{VSZ}),
while the 2nd term has been obtained from the contribution of
the {$S_2\equiv (\bar uu+\bar dd)$} scalar meson having its mass and coupling
fixed by QCD spectral sum rules \cite{SNB,SNG} and in the scheme where the
observed low mass $\sigma$ meson results from a maximal mixing between the
$S_2$ and the
$\sigma_B$ associated to the gluon component of the trace of the anomaly
\cite{VENEZIA,BRAMON,SNG} \footnote{Present data appear to favour this scheme \cite{MONTANET}.}:
\beq
\theta^\mu_\mu= \frac{1}{4}\beta(\alpha_s) G^2 +\ga
1+\gamma_m(\alpha_s)\dr\sum_{u,d,s} m_i\bar\psi_i\psi_i~,
\eeq
where $\beta$ and $\gamma_m$ are the $\beta$ function and mass anomalous
dimension. In this way, one obtains at the scale $m_c$:
\bea\label{resb6}
B_6^{1/2}(m_c)&\simeq& 3.7 \ga \frac{m_s+m_d}{m_s-m_u}\dr^2 
\times\Bigg{[}\nnb\\&& \ga
0.65\pm 0.09\dr-{(0.53\pm 0.13)}\nnb\\&& \times\ga
\frac{(m_s-m_u)~[{\rm MeV}]}{142.6}\dr\Bigg{]}~,
\eea
which satisfies the double chiral constraint \cite{REFEREE}.
We have used the running charm quark mass $m_c(m_c)=1.2\pm 0.05$ GeV \cite{SNH,JAMINC}.
Evaluating the running quark masses at 2 GeV, with the values given
in \cite{SNL}, one deduces:
\bea\label{resb6a}
B_6^{1/2}(2)&\simeq &(1.0\pm 0.4) ~{\rm for ~}m_s(2)=119~{\rm MeV},\nnb\\
&\leq&(1.5\pm 0.4) ~{\rm for ~}m_s(2)\geq 90~{\rm MeV}~.\nnb\\
\eea
The errors added quadratically have been relatively enhanced by the partial
cancellations of the two
contributions. Therefore, we deduce the combination:
\bea\label{comb}
{\cal B}_{68}&\equiv& B_6^{3/2}-0.48B_8^{3/2}\nnb\\
&\simeq& (0.3\pm 0.4)
~{\rm for}~m_s(2)=119~{\rm
MeV},\nnb\\
&\leq& (1.0\pm 0.4)
~{\rm for}~m_s(2)\geq 90~{\rm
MeV},
\eea
where we have added the errors quadratically. 
Using the approximate simplified expresssion \cite{BURAS}
\beq
\frac{\epsilon'}{\epsilon}\approx 14.5\times 10^{-4}\ga\frac{110}{\overline{m}_s(2)~[{\rm MeV}]}
\dr^2 {\cal B}_{68}~,
\eeq
one can deduce the result in units of $10^{-4}$:
\bea\label{resepsilon}
\frac{\epsilon'}{\epsilon}&\simeq& (4\pm 5)~{\rm for}~m_s(2)=119~{\rm
MeV},\nnb\\
&\leq& (22\pm 9)~,~{\rm for}~m_s(2)\geq 90~{\rm
MeV},
\eea
where the errors come mainly from ${\cal B}_{68}$ (40\%). The upper bound agrees
quite well with the world average data \cite{KTEV}:
\beq
\frac{\epsilon'}{\epsilon}\simeq (19.3\pm 2.4)\times 10^{-4}~.
\eeq
We expect that the failure of the inaccurate estimate for reproducing the data is not
na\"\i vely due to the value of the quark mass, but
may indicate the need for other important contributions than the alone
$\bar qq$ scalar meson $S_2$ (not the observed
$\sigma$)-meson which have not been considered so far in the analysis.
Among others, a
much better understanding of the effects of the gluonium (expected large
component of the $\sigma$-meson
\cite{VENEZIA,SNG,BRAMON}) in the
amplitude, through presumably a new operator needs to be studied. 
\begin{table*}
%[hbt]
\setlength{\tabcolsep}{1.pc}
%\begin{center}
\caption{ Penguin $B$--parameters for the $\Delta S=1$ process from
different approaches at $\mu=2$ GeV.
We use the value $m_s(2)=(119\pm 12)$ MeV from \cite{SNL}, and predictions
based on dispersion relations
\cite{DONO,ENJL} have been rescaled according to it. We also use for our
results $f_\pi=92.4$ MeV, but we give in the text their $m_s$ and $f_\pi$ dependences.
Results without any comments on the scheme have been
obtained in the $\overline{MS}- NDR-$scheme. However, at the present accuracy, one cannot
differentiate these results from the 
ones of $\overline{MS}- HV-$scheme.}
\begin{tabular*}{\textwidth}{@{}l@{\extracolsep{\fill}}rrrrr}
%\begin{tabular}[h]{ccccc}
\hline
%\hline
& & &&\\
{\bf Methods}&${\boldmath B^{1/2}_6}$&$B^{3/2}_8$&$B^{3/2}_7$&{\bf Comments}\\
&&&&\\
\hline
&&&&\\
Lattice \cite{MARTI}&$0.6\sim 0.8$ &$0.7\sim 1.1$ & $0.5\sim
0.8$&Huge NLO\\
&unreliable&&&at matching\cite{KILCUP}\\
&&&&\\
Large $N_c$ \cite{HAMBYE}&$0.7\sim 1.3$&$0.4\sim 0.7$&$-0.10\sim
0.04$&${\cal O}(p^0/N_c,~p^2)$\\
&&&&scheme?\\
&$1.5\sim 1.7$&$-$&$-$&${\cal O}(p^2/N_c)$; $m_q=0$\\
&&&&scheme?\\
{\bf Models}&&&\\
%&\\
Chiral QM \cite{FAB}&$1.2\sim 1.7$&$\sim 0.9$&$\approx
B^{3/2}_8$&$\mu=.8$ GeV \\
&&&&rel. with $\overline{MS}$ ?\\
ENJL+IVB \cite{PRADES}&$2.5\pm 0.4$&$1.4\pm 0.2$&$0.8\pm 0.1$&$NLO$ in $1/N_c$\\
&&&&$m_q=0$ \\
%&&&\\
%&&&&\\
L$\sigma$-model \cite{SIGMAL}&$\sim 2$&$\sim 1.2$&$-$&Not unique\\
&&&&$\mu\approx 1$ GeV; scheme ?\\
NL $\sigma$-model \cite{SIGMAN}&$1.6\sim 3.0$&$0.7\sim
0.9$&$-$&$M_\sigma$: free; $SU(3)_F$ trunc.
\\ &&&&$\mu\approx 1$ GeV; scheme ?\\
&\\
{\bf Dispersive}\\
%&\\
Large $N_c$+ LMD&$-$&$-$&0.9&$NLO$ in $1/N_c$, \\
+LSD--match.\cite{ENJL}&&&strong $\mu$-dep.&\\
&&&& \\
DMO-like SR \cite{DONO}&$-$&$1.6\pm 0.4$&$0.8\pm 0.2$&$m_q=0$\\
&&huge NLO&&Strong $s$,~ $\mu$--dep.\\ &&&\\
FSI \cite{PALLANTE}&$1.4\pm 0.3$&$0.7\pm 0.2$&$-$&Debate for fixing \\
&&&&the Slope \cite{BURAS2}\\
&&&\\
%&&&&\\
%\hline
%&&&&\\
{\bf This work}&&&&\\
%&&&&\\
DMO-like SR:&--&$2.2\pm 1.5$&$0.7\pm 0.2$&$m_q=0$\\
\cite{DONO} revisited&&inaccurate&&Strong $s$,~ $\mu$--dep.\\
&\\
$\tau$-like SR&$-$&$-$&inaccurate&$t_c$--changes\\
&&&&\\
$R^{V-A}_\tau$&$-$&$1.7\pm 0.4$&$-$&$m_q=0$\\
&&&&\\
$S_2\equiv (\bar uu+\bar dd)$&$1.0\pm 0.4$&$-$&$-$&$\overline{MS}-$scheme\\
from QSSR&$\leq 1.5\pm 0.4$&&&$m_s(2)\geq$ 90 MeV\\
&&&\\
\hline
%\hline
\end{tabular*}
%\end{center}
\end{table*}
\nin
\section{Summary and conclusions}
We have explored the $V$-$A$ component of the hadronic tau decays for predicting non-perturbative
QCD parameters. Our main results are summarized as:
\begin{itemize}
\item QCD contiuum threshold - transition between the low- and high-energy regimes:
Eq. (\ref{tc}).
\item Low-energy constants $L_{10}$, $m_{\pi^\pm}-m_{\pi^0}$ and $f_\pi$ in the chiral limit:
\begin{itemize}
\item
Eq. (\ref{pred}) (best)
\item Eq (\ref{conserve}) (conservative).
\end{itemize}
\item Electroweak penguins: 
\begin{itemize}
\item Eq. (\ref{b78}): $B^{3/2}_7$, 
\item Eq.
(\ref{resb8}): $B^{3/2}_8$\item Eq.
(\ref{resq8}): $\la (\pi\pi)_{I=2}|{\cal Q}_{8}^{3/2}|K^0\ra$.
\end{itemize}
\item $\epsilon'/\epsilon$: Eq. (\ref{resepsilon}) 
\end{itemize}
Our results are compared with some other predictions in Table 1. However, as mentioned in the
table caption, a direct comparison of these results is not straightforward due to the different
schemes and values of the scale where the results have been obtained. In most of the approaches,
the values of $B^{3/2}_7$ are in agreement within the errors and are safely in the range $0.5\sim
1.0$. For $B^{3/2}_8$ the predictions can differ by a factor 2 and cover the range $0.7 \sim 2.1$.
There are strong disagreements by a factor 4 for the values of $B^{1/2}_6$ which range from
$0.6\sim 3.0$. We are still far from a good control of these non-perturbative parameters, which
do not permit us to give a reliable prediction of the $CP$ violation parameter
$\epsilon'/\epsilon$. Therefore, no definite bound for new physics effects can be derived at
present, before improvements of these ESM predictions.   
\vfill\eject

\end{document}